\documentclass[12pt]{iopart}

\usepackage{graphicx}  

\begin{document}
\title{On survey of nuclei and hypernuclei in multifragmentation}

\author{N. Buyukcizmeci$^{1}$, R. Ogul$^{1}$, A.S. Botvina$^{2,3}$, M. Bleicher$^{2,4,5,6}$}
\address{$^{1}$Department of Physics, University of Sel\c{c}uk, 42079
Konya, Turkey.} 
\address{$^{2}$Institute for Theoretical Physics, J.W. Goethe University, D-60438 Frankfurt am Main, Germany.} 
\address{$^{3}$Institute for Nuclear Research, Russian Academy of Sciences, 117312 Moscow, Russia.}
\address{$^{4}$GSI Helmholtz Center for Heavy Ion Research, 64291 Darmstadt, Germany.}
\address{$^{5}$John-von-Neumann Institute for Computing (NIC), FZ Julich, Julich, Germany.}
\address{$^{6}$Helmholtz Forschung Akademie Hessen (HFHF), Frankfurt, Germany.}

\ead{ nihal@selcuk.edu.tr}
\vspace{10pt}
\begin{indented}
\item[] 10 March 2020
\end{indented}

\begin{abstract}
Multifragmentation reactions are dominating processes for the decomposition of highly excited nuclei leading to the fragment production in heavy-ion collisions. At high energy reactions strange particles are abundantly produced. We present a novel development of the Statistical multifragmentation model (SMM) as its generalization for the hyper-matter which is formed after the hyperon capture. In this way, it is possible to describe its disintegration into normal and hyper-nuclei. Some properties of hyper-nuclei and their binding energies can be determined from the comparison of the isotope yields. The main focus of this method is to investigate strange and multi-strange hypernuclei since their properties are not easy to measure in traditional hyper-nuclei experiments.
\end{abstract}

%
% Uncomment for keywords
%\vspace{2pc}
\noindent{\it Keywords}:mass yield, isotope distributions, binding energy of hypernuclei, multifragmentation
%
% Uncomment for Submitted to journal title message
\submitto{\PS}
%
% Uncomment if a separate title page is required
%\maketitle
% 
% For two-column output uncomment the next line and choose [10pt] rather than [12pt] in the \documentclass declaration
%\ioptwocol
%

\section{Introduction}

Different hyperons, such as $\Lambda$, $\Sigma$, $\Xi$ and $\Omega$, can be produced in relativistic nuclear and ion collisions. In peripheral reactions, these hyperons may be captured by the projectile and target residues, and hypernuclei are formed. In this case, one can investigate the hypernuclei formation, the hyperon--nucleon and the hyperon--hyperon interactions. This kind of investigations are crucial for hypernuclear physics, traditional nuclear physics, particle physics and nuclear astrophysics \cite{Ban90,Has06,Sch93,Gal12,Buy13,Hel14}. In these reactions we may also expect modifications of properties of strange particles in nuclear matter, that may influence their formation, capture, and life-time of hypernuclei. The knowledge of modified properties are important for the progress in QCD theory and for astrophysical applications of hypernuclear physics in neutron stars. For this purpose, one may include the investigation of particle correlations and hypernuclei with exotic isospin content and multi-hyperon ones, since they provide a guide to novel states of matter. This kind of studies are important for the future experiments to search for the hypernuclei and to address new fragmentation reactions for production of hypernuclei.

The hypernuclei were discovered in the deep-inelastic reactions leading to fragmentation processes, long ago \cite{Dan53}. Many experimental collaborations focused on investigation of hypernuclei and their properties in the reactions induced by relativistic particles (STAR \cite{star}, ALICE \cite{alice}, PANDA \cite{panda}, CBM \cite{Vas17}, HypHI, Super-FRS and R3B\cite{saito-new,super-frs}, BM@N and MPD  \cite{nica}). Hyperon production and their capture in transport models at sub-threshold and high energies was investigated by using the microscopic transport models UrQMD, DCM, GiBUU, and PHSD \cite{Bot15,Jan16,Bot11,Bot17,giessen}. Besides theoretical studies \cite{Buy13,Bot15,Bot11,Bot17,giessen,Bot07,Bot13,Cas95} about strange and multi-strange hypernuclei formed in nuclear collisions, there also exist experimental confirmations of the formation of hypernuclei \cite{saito-new,Arm93,Ohm97}.  
We have presently focused on the theoretical research of the fragmentation reactions regarding $\Lambda$ hypernuclei production. These hypernuclei can be beyond the neutron and proton drip lines \cite{Buy13,Sam17}. In our previous studies \cite{Buy13,Bot07,Buy18,Bot16}, we developed hyper-SMM code for different processes, such as fragment formation mechanisms including the evaporation and fission of hyper-nuclei at low energies. We have shown a new double ratio method to determine the hyperon binding energies for strange and in multi-strange nuclei by using the analysis of the relative yields of hypernuclei \cite{Buy18}. In this paper, we will give further details of our research, and compare our  results for different initial sized sources to show the size effect, for a complementary view.

\section{Hypernuclei in multifragmentation regime}

As it was reported in Refs. \cite{Bot15,Bot13,Bot17}, the yields of the hyperresidue nuclei in 
peripheral ion collisions may saturate around 3--5 A GeV of energies in the laboratories. In these deep-inelastic processes, it is possible to have large hyper-residues with broad ranged mass number and excitation energy \cite{Bot15,Bot11,Bot17}. 
In particular, the saturation yields are large enough for single (100 mb), double (1 mb) and triple  (0.01 mb) hypernuclear residues in Pb+Pb collisions \cite{Bot17}. One may use these yields to estimate extension of the hypernuclei chart and for hypernuclear investigations. 

The masses, excitation energies and formation of excited nuclear residues in nucleus-nucleus and hadron-nucleus collisions at high-energies were analyzed within fragmentation and multifragmentation reactions, both experimentally and theoretically \cite{Bot17,Xi97}. At high excitation energies, multifragmentation 
process would be the dominant channel \cite{SMM,Poc97,EOS}. As is well known, multifragmentation is a fast process (approximately 100 fm/c), for which the chemical equilibration should be reached in the reaction system. In the freeze-out volume, the strong interactions are assumed to exist among the baryons located in the neighbourhood of each other. Since the statistical model predictions are in well agreement with the fragmentation and multifragmentation data \cite{Xi97,SMM,EOS,Ogu11,Bot95,Ergun15,Imal15}, this statistical approach was extended to hypernuclear systems in Ref. \cite{Bot07}. 
%%%%%%%%%%%%%%%%%%%%%%%%%%%%%%%%%%%%%%%%%%%%%%%%%%%%%%%%%%%%%%%%%%%%%%%%
\begin{figure}[htbp]
\centering
\includegraphics[width=9cm,height=16cm]{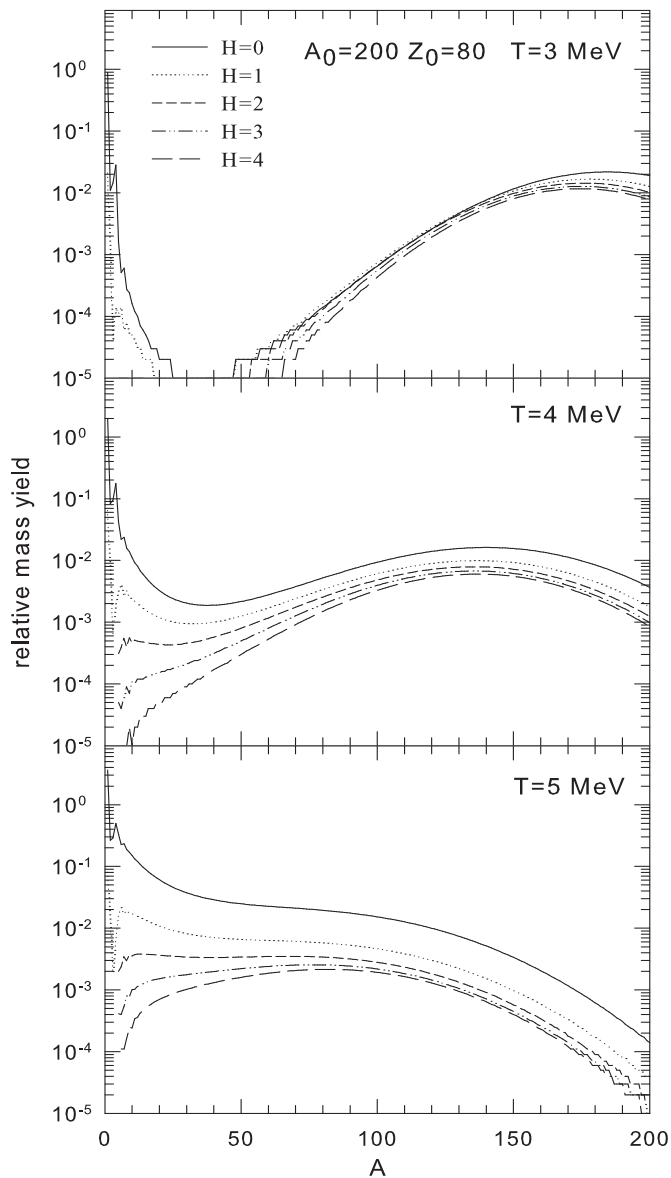}
\caption{\small{Relative yields of fragments and hyperfragments with H hyperons
versus their mass number, after decay of excited nucleus, with $A_0$=200, $Z_0$=80,
containing $H_0$=4 $\Lambda$ hyperons at $T=3,$ 4, and $5$ MeV.
Calculations are performed in grand canonical approximation.}}
\label{fig1}
\end{figure}
%%%%%%%%%%%%%%%%%%%%%%%%%%%%%%%%%%%%%%%%%%%%%%%%%%%%%%%%%%%%%%%%%%%%%%%%
%%%%%%%%%%%%%%%%%%%%%%%%%%%%%%%%%%%%%%%%%%%%%%%%%%%%%%%%%%%%%%%%%%%%%%%%
\begin{figure}[htbp]
\centering
\includegraphics[width=8cm,height=12cm]{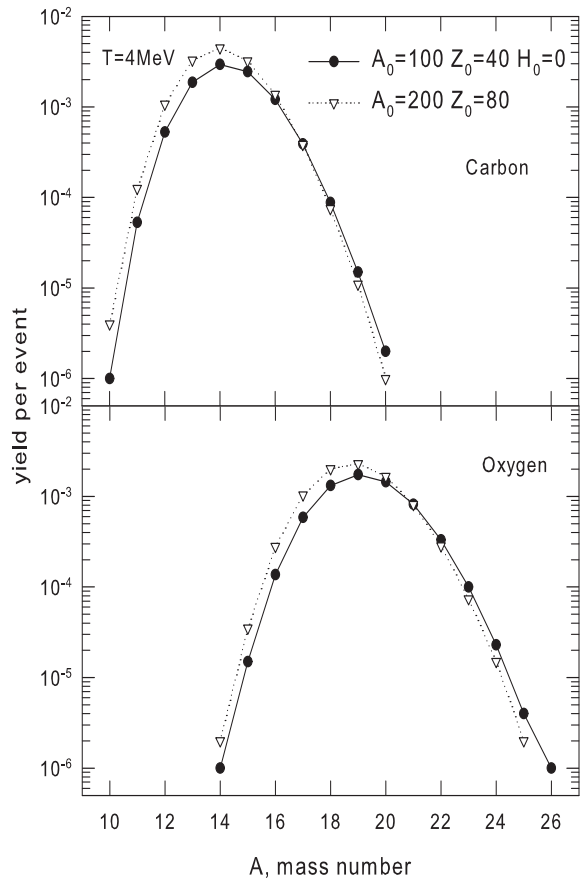}
\caption{\small{Relative yields of Carbon (top panel) and Oxygen isotopes (bottom panel) as a function of their mass number, 
after decay of excited initial nuclei with $A_0$=100, $Z_0$=40 (full symbols)
and $A_0$=200, $Z_0$=80 (open symbols), at $T=4$ MeV.}}
\label{fig2}
\end{figure}
%%%%%%%%%%%%%%%%%%%%%%%%%%%%%%%%%%%%%%%%%%%%%%%%%%%%%%%%%%%%%%%%%%%%%%%%
The mean value of the yields for individual fragments with the mass (baryon) number $A$, the charge number $Z$, and the $\Lambda$ hyperon number $H$ is defined in the grand canonical approximation as follows: 
\begin{eqnarray} \label{yazh} 
Y_{\rm A,Z,H}=g_{\rm A,Z,H}\cdot V_f\frac{A^{3/2}}{\lambda_T^3} 
{\rm exp}\left[-\frac{1}{T}\left(F_{A,Z,H}-\mu_{AZH}\right)\right], 
\nonumber\\ 
%\mu_{AZH}=A\mu+Z\nu+H\xi~, 
\mu_{AZH}=A\mu+Z\nu+H\xi .~~~~
\end{eqnarray} 
Here, $T$, $F_{A,Z,H}$, $V_f$, $g_{\rm A,Z,H}$, $\lambda_T=\left(2\pi\hbar^2/m_NT\right)^{1/2}$, and $m_N$ represent the temperature, the internal free energies of the fragments, the free available volume for the translational motion of the fragments,the spin degeneracy factor of species $(A,Z,H)$, the baryon thermal wavelength, and the average 
baryon mass, respectively. Taking into account the conservation of the total mass (baryon) number $A_0$, charge number $Z_0$, and $\Lambda$ hyperon number $H_0$ in the system, the corresponding chemical potentials $\mu$, $\nu$, and $\xi$ can be numerically calculated.
The statistical ensemble take into accounts all break-up channels of baryons and excited 
fragments. As the hot fragments are formed in the freeze-out volume $V$, 
the excluded volume approximation can be written as $V=V_0+V_f$, where 
$V_0=A_0/\rho_0$ ($\rho_0\approx$0.15 fm$^{-3}$ is the normal nuclear 
density), and the free volume $V_f=\kappa V_0$, with 
$\kappa \approx 2$, as shown in Refs. \cite{Xi97,EOS,Ogu11}.

Nuclear clusters are defined on the basis of the liquid-drop approximation: 
For the particles with $A < 4$ (nuclear gas), the experimental binding energies were treated as in Refs. \cite{Ban90,Has06,SMM}. For the case of the fragments with $A=4$ the excitation
energy is written as $E_{x}=AT^{2}/\varepsilon_0$ for corresponding excited states of $^{4}$He,
$^{4}_{\Lambda}$H, and $^{4}_{\Lambda}$He, where $\varepsilon_0
\approx$16 MeV is the inverse volume level density parameter Ref. \cite{SMM}.  Larger fragments with 
$A > 4$ are assumed to be liquid dropes. $g_{\rm A,Z,H} = 1$ is taken for the fragments with 
$A > 4$, since they are excited and do populate many states above the ground state according to the given 
temperature dependence of the free energy. Nuclear liquid-gas coexistence in the freeze-out volume is valid also for hypermatter. The internal free energies of nuclei and hypernuclei are written as follows: 
\begin{equation}  \label{fld}
F_{A,Z,H}=F_{A}^B+F_{A}^S+F_{AZH}^{\rm sym}+F_{AZ}^C+F_{AH}^{\rm hyp}~~.
\end{equation}
Here, $F_{A}^B$ represents the bulk energy, $F_{A}^S$ the surface energy, $F_{AZH}^{\rm sym}$ the symmetry energy, and $F_{AZ}^C$ the Coulomb energy \cite{SMM}. The free hyper energy term $F_{AH}^{\rm hyp}$ is determined only by the binding energy of hyper-fragments. The knowledge about the experimentally established masses of single light hypernuclei
\cite{Ban90,Has06} is very limited in the literature, and there exist only few measurements for the double hypernuclei.
In Ref. \cite{Bot07} we suggested a liquid drop hyper term as follows:
\begin{equation} \label{ht}
F_{AH}^{\rm hyp}=(H/A)\cdot(-10.68 A + 21.27 A^{2/3})~MeV .
\end{equation}
Here, ($H/A$) is the fraction of $\Lambda$ hyperons. For the liquid-drop parametrization, we consider only the volume contribution and the surface term, in case of the saturation of the nuclear interactions.

%%%%%%%%%%%%%%%%%%%%%%%%%%%%%%%%%%%%%%%%%%%%%%%%%%%%%%%%%%%%%%%%%%%%%%%%
\begin{figure}[htbp]
\centering
\includegraphics[width=8cm,height=12cm]{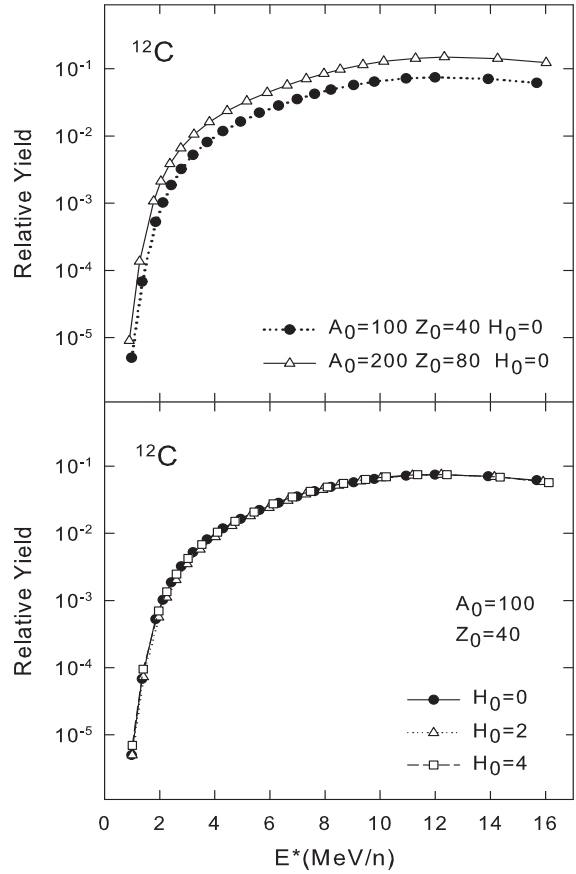}
\caption{\small{Top panel shows relative yields of Carbon isotopes as a function of the excitation energy, 
after decay of excited initial nuclei with $A_0$=100, $Z_0$=40 (full circles), for $H_0$=0,
and $A_0$=200, $Z_0$=80 (open triangles). Bottom panel demonstrate the results for $A_0$=100, $Z_0$=40, but for different values of $H_0$=0, 2, 4.}}
\label{fig3}
\end{figure}
%%%%%%%%%%%%%%%%%%%%%%%%%%%%%%%%%%%%%%%%%%%%%%%%%%%%%%%%%%%%%%%%%%%%%%%%
%%%%%%%%%%%%%%%%%%%%%%%%%%%%%%%%%%%%%%%%%%%%%%%%%%%%%%%%%%%%%%%%%%%%%%%%
\begin{figure}
\begin{center}
\includegraphics[width=8cm,height=12cm]{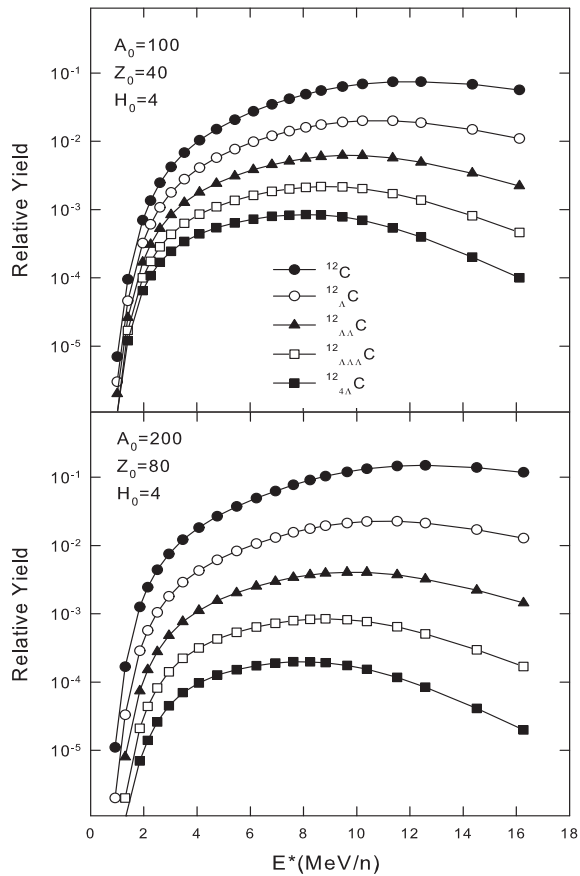}
\end{center}
\caption{\small{
Isotope yields of Carbon isotopes produced after decay of 
an excited hypernuclear system at different excitation energies.
Top panel shows the results for $A_0$=100, $Z_0$=40, and bottom panel for $A_0$=200, $Z_0$=80, with initial $\Lambda$ hyperon number $H_0$=1, 2, and 4. 
}}
\label{fig4}
\end{figure}
%%%%%%%%%%%%%%%%%%%%%%%%%%%%%%%%%%%%%%%%%%%%%%%%%%%%%%%%%%%%%%%%%%%%%%%%
%%%%%%%%%%%%%%%%%%%%%%%%%%%%%%%%%%%%%%%%%%%%%%%%%%%%%%%%%%%%%%%%%%%%%%%%
\begin{figure}[htbp]
\begin{center}
\includegraphics[width=8cm,height=12cm]{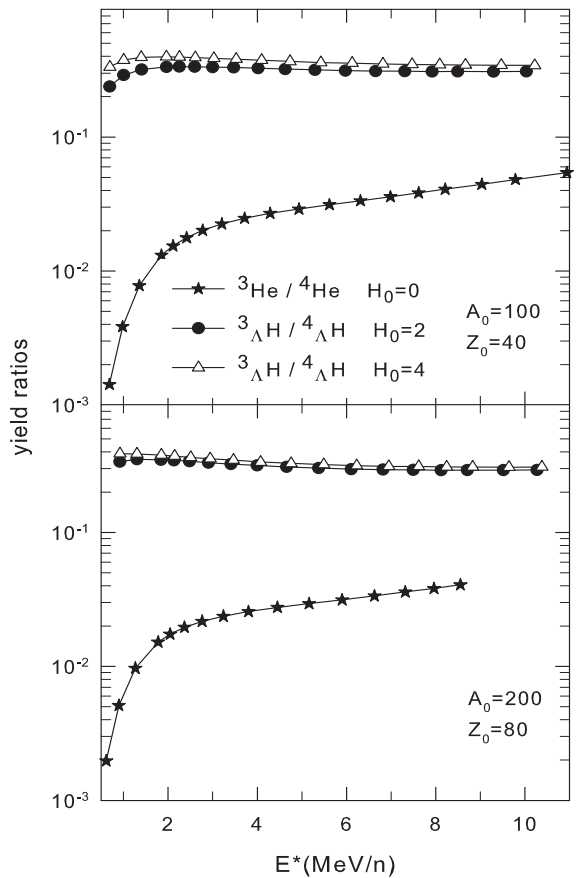}
\end{center}
\caption{\small{Yield ratios of the produced fragments, $^{3}$He/$^{4}$He ($H_0$=0) and $^{3}$H$_{\Lambda}$/$^{4}$H$_{\Lambda}$ ($H_0$=2 and 4) as a function of the excitation energy, for initial nuclei $A_0$=100, $Z_0$=40 (top panel) and $A_0$=200, $Z_0$=80 (bottom panel).
}}
\label{fig5}
\end{figure}
%%%%%%%%%%%%%%%%%%%%%%%%%%%%%%%%%%%%%%%%%%%%%%%%%%%%%%%%%%%%%%%%%%%%%%%%
%%%%%%%%%%%%%%%%%%%%%%%%%%%%%%%%%%%%%%%%%%%%%%%%%%%%%%%%%%%%%%%%%%%%%%%%
\begin{figure}[htbp]
\begin{center}
\includegraphics[width=8cm,height=12cm]{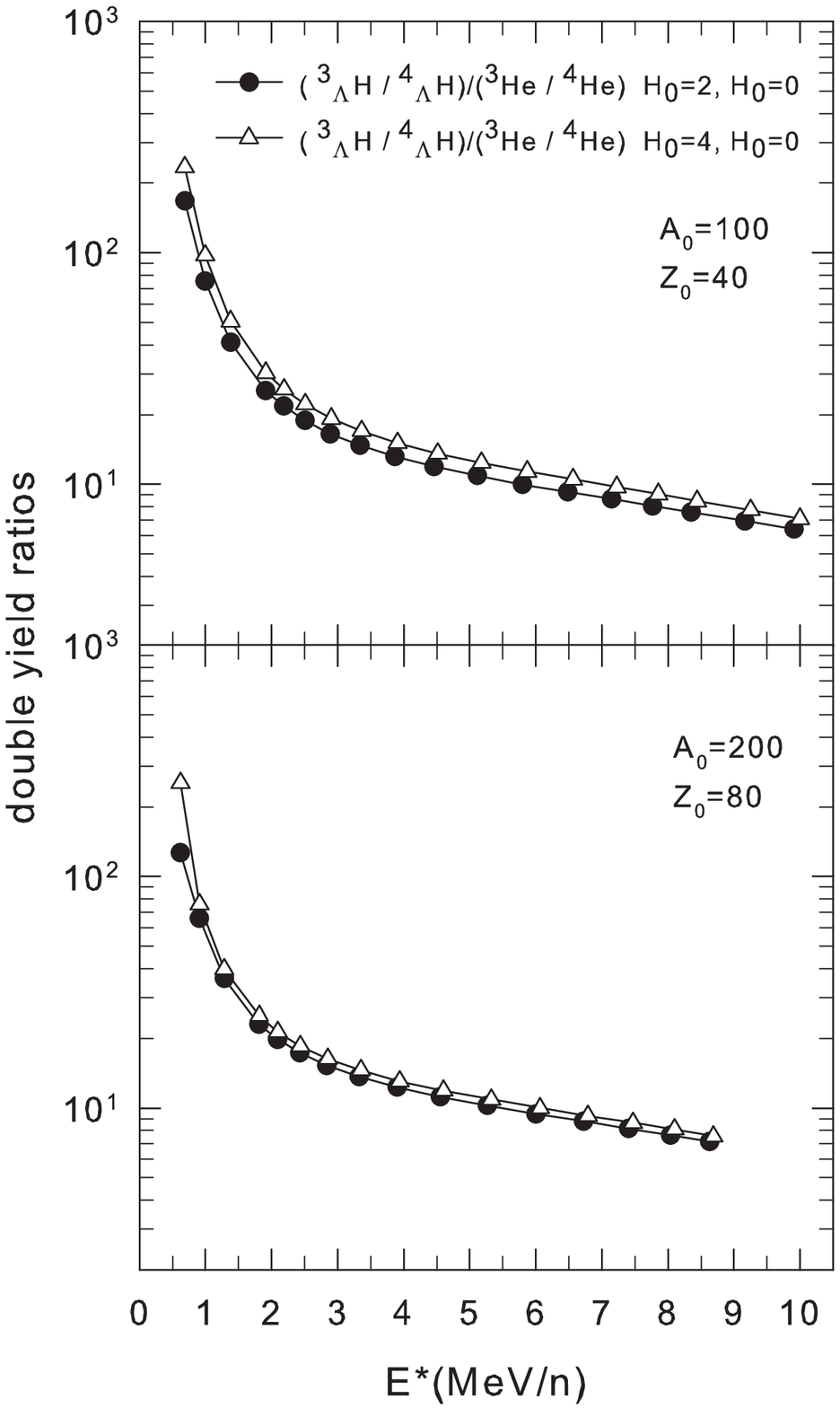}
\end{center}
\caption{\small{Double yield ratios of the produced fragments ($^{3}$H/$^{4}$H)/($^{3}$He$_{\Lambda}$/$^{4}$He$_{\Lambda}$),  as a function of the excitation energy, for initial nuclei with $A_0$=100, $Z_0$=40 (top panel), and $A_0$=200, $Z_0$=80 (bottom panel), and for the hyperon numbers $H_0$=0, 2 and 4.
}}
\label{fig6}
\end{figure}
%%%%%%%%%%%%%%%%%%%%%%%%%%%%%%%%%%%%%%%%%%%%%%%%%%%%%%%%%%%%%%%%%%%%%%%%

In Fig.~1, we have demonstrated the total mass distributions of hot fragments produced after the decay of hot excited residue with mass number $A_0$=200 and charge number $Z_0$=80, for initial $\Lambda$ hyperon number $H_0$=4 at different temperatures $T$=3, 4, and 5 MeV. Grand canonical ensemble calculations are permorfed to show the evaluation of mass distributions. While the normal hot nuclei distributions have highest probability, total distributions with additional $\Lambda$ hyperon have lower values depending on the number of $\Lambda$ hyperons. There is an ordering in distributions for all panels of Fig.1. This is because hypernuclear residue system will have higher binding energies with extra $\Lambda$ hyperons inside, and more energy will be required to decay for such hypernuclei. Even if the hyperon-nucleon potential of hypernuclei is close to nucleon-nucleon potential, normal hot fragments would have priority to decay before hyperfragments.  As can be seen from Fig. 1, clearly, while larger fragments are dominant at $T$=3 MeV, the differences between distributions for different $\Lambda$ hyperon contents have considerably close values for $A>100$.  In the middle panel, relative mass distributions show plateau-like behaviour around $T$=4 MeV, and the number of intermediate mass fragments increases as expected for intermediate energies. In the bottom panel, flat distribution starts to change like exponential fall off, with increasing temperatures. 

Fragment yields of C and O isotopes emitted from the initial excited nuclei with $A_0$=100, $Z_0$=40, and $A_0$=200, $Z_0$=80 around at $T$=4 MeV, are shown in Fig. 2. In Fig.~3, we show the results for relative yields of C isotopes as a function of the excitation energy. We also show the relative yields of C isotopes emitted from the initial nuclei with $A_0$=100, $Z_0$=40, and $A_0$=200, $Z_0$=80 (top panel) and comparison of the yields  for different $H_0$=0, 2, and 4 values for $\Lambda$ captured hyperons for $A_0$=100, $Z_0$=40 system (bottom panel). Thus, one can see the size effect on the yields from the top panel of this figure. However, the relative yield values of C are close to each other for initial $A_0$=100, $Z_0$=40 and $H_0$=0, 2, and 4 system as shown in the bottom panel.  This trend is also the same for initial nucleus $A_0$=200, $Z_0$=80 and $H_0$=0, 2, and 4. It is also seen that big initial system produces the yields with higher probability, as is expected.

In Fig. 4, we present the relative yields of $^{12}C$, $^{12}_{\Lambda}C$, $^{12}_{\Lambda\Lambda}C$, $^{12}_{\Lambda\Lambda\Lambda}C$, and $^{12}_{4\Lambda}C$ as function of excitation energies for both initial stems. These isotopes are emitted from the excited sources with ($A_0$=100, $Z_0$=40) and ($A_0$=200, $Z_0$=80) for $H_0$=4 nuclei. From Fig. 4, one may observe ordered distributions of the yields from normal nuclei with higher probability to hypernuclei. While the relative yields of normal and hyper-Carbon isotopes show an increasing trend up to 3-4 MeV/nucleon, they show almost a flat distribution after 5-6 MeV/nucleon. Since the envelope line of all isotopes distributions (versus A) corresponds to mass distributions, this behaviour supports the trend of mass distributions of hot fragments in Fig. 1.

In Fig. 5, the yield ratios of $^{3}He/^{4}He$ and $^{3}_{\Lambda}H/^{4}_{\Lambda}H$ are demonstrated for the both initial systems. The foreseen  experimental technology at FAIR is expected to give opportunity to measure these ratios for $^{3}_{\Lambda}H/^{4}_{\Lambda}H$. The ratio seems to produce almost a flat distribution with increasing excitation energies. Furthermore, one can investigate the amount of produced yields and their relations, by taking into account the double ratios of  $^{3}_{\Lambda}H/^{4}_{\Lambda}H$ and $^{3}He/^{4}He$, as shown in Fig. 6. In all cases, the additional $\Lambda$ hyperon reflects its effect inside the nucleus, for the medium. That is why it is very important to have the knowledge of the binding energies of hypernuclei. 

\section{Determination of the hyperon binding energies}

Our method is independent from the theoretical assumptions on the hypernuclei masses, but it relies on the fragment yields obtained in the multifragmentation reactions \cite{a2017}. Primary and cold fragments can be found in the freeze-out volume, and these possibilities are strictly related with the production mechanisms. It is possible to determine the evolution of the fragment production for selected reaction events in  multifragmentation experiments. Then, the hypernuclei yields can be used to determine the hyperon binding energies \cite{Buy18}. 

Here, we suggest a method to determine the binding energy of a hyperon inside the nucleus. Let us consider two hyper-nuclei 
with different masses ($A_1,Z_1,H$) and ($A_2,Z_2,H$), 
while the difference of the number of $\Lambda$ hyperons is one. 
First we take the double ratio 
of $Y_{\rm A_1,Z_1,H}/Y_{\rm A_1-1,Z_1,H-1}$ to 
$Y_{\rm A_2,Z_2,H}/Y_{\rm A_2-1,Z_2,H-1}$. 
Then, taking the logarithm of the double ratio, the hypernuclei yields and the hyperon binding energies can be written as
\begin{equation} \label{DRL} 
\Delta E_{A_1A_2}^{\rm bh} = T \cdot \left[ ln(\frac{C_{\rm A_1,Z_1,H}}{ C_{\rm A_2,Z_2,H}})- 
ln(\frac{Y_{\rm A_1,Z_1,H}/Y_{\rm A_1-1,Z_1,H-1}}{Y_{\rm A_2,Z_2,H}/Y_{\rm A_2-1,Z_2,H-1}}) \right] .
\end{equation} 
Here, we have \begin{equation} \label{dEbh} 
\Delta E_{A_1A_2}^{\rm bh}=  E_{A_1}^{\rm bh}-E_{A_2}^{\rm bh}.
\end{equation} 
The term $C_{\rm A,Z,H}=(g_{\rm A,Z,H}/g_{\rm A-1,Z,H-1}) \cdot (A^{3/2}/(A-1)^{3/2})$ 
depends on the ratio of the spin factors of ($A_1,Z_1,H$),($A_1-1,Z_1,H-1$), ($A_2,Z_2,H$), and 
($A_2-1,Z_2,H-1$) nuclei, and very weakly on $A_1$ and $A_2$.
As is seen from these equations, the temperature of the  system and the difference between the hyperon separation energies of the fragments are related to each other through the double yield ratio method. 

%\subsection{Numerical result of the method}

We have shown that the numerical results can be obtained by using double ratio approach for hypernuclei with any number of 
hyperons. The equation (\ref{DRL}) is valid for $H \geq 1$. In heavy-ion nuclear reactions one can 
obtain multi-strange residues with a quite large 
probability \cite{Bot17}, and a very wide mass/isospin range that will be 
available for examination. Consequently, it is possible to get direct experimental 
evidences for hyperon binding energies in double hypernuclei and multi-hyperon 
nuclear matter, after the normalization of the binding energy of one known hypernucleus $A_2$, by using the relative hyperon binding energies $\Delta E_{A_1A_2}^{\rm bh}$ and its absolute values. 

In Fig.~7, we show the differences in hyperon binding energies 
$\Delta E_{A_1A_2}^{\rm bh}$ (the notation is shortened to 
$\Delta E_{\rm bh}$) as function of the mass number 
difference $\Delta A = A_2 - A_1$ of isotopes. In the top panel, we show the results for single hypernuclei and normal 
nuclei with $A_0$=200, $Z_0$=80, and 
hyperon number $H_0$=4. Bottom panel shows the results for $H_0$=3 obtained by using hyper-SMM code given in Refs. (Refs. \cite{Buy13,Bot07}). 
In the top panel, we show the double ratios of isotopes for hyper-$^{13}$C nucleus with $A_1$ and $A_2$=21, 25, 33, 41, 50, 60, 81, 101, 125 are assumed to get a broad range of $\Delta A$. 
We take the yield ratios of 
$^{21}_{\Lambda}$O/$^{20}$O, 
$^{25}_{\Lambda}$Mg/$^{24}$Mg, 
$^{33}_{\Lambda}$P/$^{32}$P, 
$^{41}_{\Lambda}$S/$^{40}$S, 
$^{50}_{\Lambda}$Ca/$^{49}$Ca, 
$^{60}_{\Lambda}$Cr/$^{59}$Cr, 
$^{81}_{\Lambda}$Ge/$^{80}$Ge, 
$^{101}_{\Lambda}$Zr/$^{100}$Zr, 
$^{125}_{\Lambda}$Sn/$^{124}$Sn, 
to $^{13}_{\Lambda}$C/$^{12}$C.
Double ratios of isotopes for hyper-$^{25}$Mg nucleus with
$A_1$ and $A_2$=33, 41, 50, 60, 81, 101, 125 are 
selected similarly as in the top figure. For simplicity, we show initial ratios as in the top figure $IR=^{13}_{\Lambda}$C/$^{12}$C
and in the bottom figure $IR=^{25}_{\Lambda}$Mg/$^{24}$Mg.
We investigate the sensitivity of our results for hot excited system by assuming the temperature $T$=4 MeV as
in fragmentation and multifragmentation reactions. 
It is shown that the extracted $\Delta E_{\rm bh}$ values increase, regularly, 
with $\Delta A$. As we have demonstrated in Ref. \cite{Buy18}, it is interesting that,
variation of $\Delta E_{\rm bh}$ with $\Delta A$ values follows the same trend for both single and double hypernuclei. 

%%%%%%%%%%%%%%%%%%%%%%%%%%%%%%%%%%%%%%%%%%%%%%%%%%%%%%%%%%%%%%%%%%%%%%%%
\begin{figure}[htbp]
\begin{center}
\includegraphics[width=8cm,height=12cm]{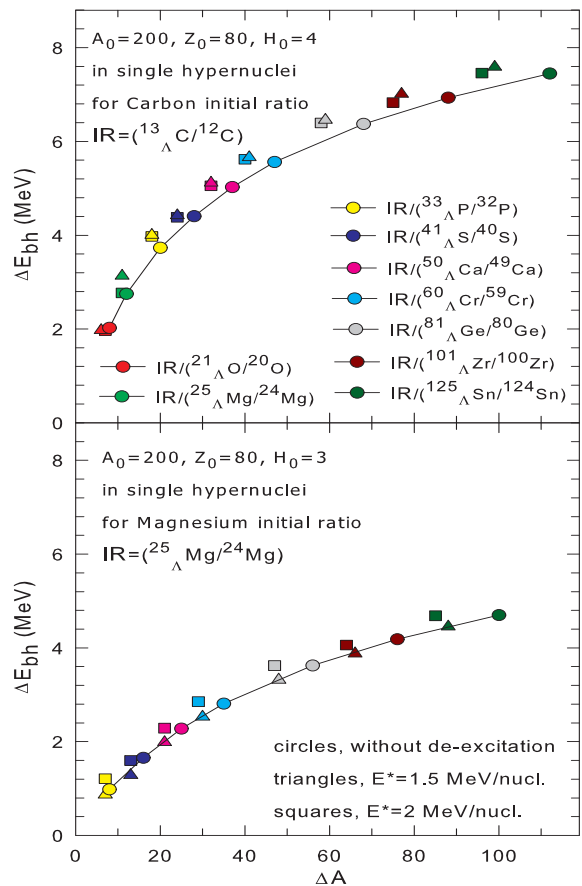}
\end{center}
\caption{\small{ (Color online)
Hyperon binding energies $\Delta E_{\rm bh}$ calculated for single hypernuclei, as a function of the difference of mass numbers $\Delta A$. The circle symbols with solid lines are for hot fragments (at $T$=4 MeV), triangles for cold fragments at $E^*$=1.5 MeV/nucleon, and squares for cold fragments at $E^*$=2.0 MeV/nucleon. Hot mother and her cold daughter nuclei have the same color symbols, and $IR$ denotes the initial ratio.}}
\label{fig7}
\end{figure}
%%%%%%%%%%%%%%%%%%%%%%%%%%%%%%%%%%%%%%%%%%%%%%%%%%%%%%%%%%%%%%%%%%%%%%%%

We also show in Fig. 7, the effect of the secondary 
de-excitation on the results for $\Delta E_{\rm bh}$ for single hypernuclei. The same mother nuclei at the temperature $T$=4 MeV are taken for initial fragment yields to be consistent with liquid phase transition region in multifragmentation reactions. In order to extract $\Delta E_{\rm bh}$ values from the experiments within the double ratio approach, we assume the temperature $T$=4 MeV for the decay
of hypernuclear system taking into account the intensive investigations 
in connection with multi-fragment formation: using kinetic energies of fragments, 
excited states population, and isotope thermometers \cite{Poc97,Bon98,Kel06}. 
In the caloric curve, plateau region is compatible within the temperatures $T=4-5$ MeV, and the excitation energy  $E*>2-3$ MeV/nucleon \cite{SMM,Poc97}. 
As is reported in Refs. \cite{Buy13,Bot07,Buy18}, 
a small number of hyperons admission in the system does not change the caloric 
curve, while the temperature becomes slightly lower. The daughter nuclei with 
the largest probability yield, selected to calculate the double ratios 
after the evaporation by taking inco account the same daughter nuclei, 
can be produced after evaporation of other nuclei with the neighbour $A$ and $Z$.
The weight of all primary nuclei after multifragmentation in the freeze-out 
volume and evaluated their contribution in the yields of daughter ones 
are also considered. The values of $\Delta E_{\rm bh}$ are determined according to the formula (\ref{DRL}). 
After evaporation as shown by triangle and square symbols 
for $E^*$=1.5 and 2 MeV/nucleon in Fig. 7, there is a regular behaviour of 
$\Delta A$ values turning to a lower ones as a result of nucleon losses. 
This trend can be seen by comparing the circles for hot primary 
fragments, with the triangles and squares for cold fragments, 
more considerably for big nuclei since the excitation energies 
of big nuclei are higher. From the comparison of the top and bottom panels of Fig. 7, one can clearly see that the binding energies can be reproduced by the different initial mother nuclei and $IR$ values for different mass intervals of nuclei.

\section{Conclusion}

Fragmentation reactions with relativistic ions offer new opportunities for production and investigation of hypernuclei. We expect production of many different hypernuclei with various masses and isospin content. We demonstrated that binding energies of strange nuclei, initially for $\Lambda$ hypernuclei, can be surely extracted by using the yields of produced fragments in our method. In future experiments, if it would be possible to measure at least a couple of hypernuclei in one experiment, this method can be used to extrapolate the binding energies of unknown hypernuclei. We hopefully wait for the results of the experiments to be performed with new technological setups, for the fragment detection, soon \cite{aumann,frs,aysto}. Thus, the hypernuclei with sufficiently large cross sections can be measured in the projectile and target spectator regions, under chemical equilibrium conditions, in the peripheral relativistic ion-ion and hadron-ion deep-inelastic collisions. The double ratio method can be used to determine their binding energies. Important task is the identification of hypernuclei, as known 
in recent ion experiments of searching for strange particles. Even if it is not easy to measure 
the hypernuclei binding energy for neutron-rich/poor nuclear species within traditional hypernuclear 
experiments, the experimental extraction of the difference of the hyperon binding energies between 
hypernuclei ($\Delta E_{A_1A_2}^{\rm bh}$) via their yields would be an alternative way. One can consider the pairs of large hyper isotopes, and afterwards of their weak decay, one can determine products far from the reaction center by using the vertex technique. 
We have also demonstrated how the mass and isotopic distributions of the products of excited hot residue nuclei evolve with temperature and excitation energies. As established during previous multifragmentation studies \cite{Xi97,SMM}, 
the temperature of the nuclear fragmenting system can be determined by selecting special observables in experiment.  
The statistical approach to describe yields of nuclei and hypernuclei implicates that the double ratio method presents an open window to understand the hypernuclear chart. We believe that our theoretical findings would be realized in future experiments at intermediate energies, at FAIR (GSI) and NICA (JINR).

\section*{Acknowledgements}
This study is supported by Scientific and 
Technological Research Council of Turkey (TUBITAK), 
under Project No. 118F111, and has been performed 
in the framework of COST Action CA15213 THOR. 
A.S. Botvina acknowledges the support of BMBF (Germany). 
We thank Frankfurt Institute for Advance Studies (FIAS) 
and Helmholtz Research Academy Hessen for FAIR (HFHF) 
in Frankfurt for kind hospitality during research visit.

\end{document}